\documentclass[12pt]{article}%
\usepackage{amsfonts}
\usepackage{cite}
\usepackage{amsmath}
\usepackage{amssymb}
\usepackage{graphicx}%
\providecommand{\U}[1]{\protect\rule{.1in}{.1in}}
\makeatletter
\@addtoreset{equation}{section}
\makeatother

\begin{document}
\begin{titlepage}
\begin{center}
\renewcommand{\thefootnote}{\fnsymbol{footnote}}
{\Large{\bf Phantom Metrics With Killing Spinors}}
\vskip1cm
\vskip 1.3cm
W. A. Sabra
\vskip 1cm
{\small{\it
Centre for Advanced Mathematical Sciences and Physics Department\\
American University of Beirut\\ Lebanon \\}}
\end{center}
\bigskip
\begin{center}
{\bf Abstract}
\end{center}
We study metric solutions of Einstein-anti-Maxwell theory admitting Killing spinors.
The analogue of the IWP metric which admits a space-like Killing vector is found and is
expressed in terms of a complex function satisfying the wave equation in flat (2+1)-dimensional space-time.
As examples, electric and magnetic Kasner spaces are constructed by allowing the solution to depend only on the time coordinate.
Euclidean solutions are also presented.
\end{titlepage}

\section{Introduction}

The first systematic classification of metrics admitting supercovariantly
constant spinors in Einstein-Maxwell theory was performed many years ago by
Tod in \cite{tod}. The analysis of Tod was to some extent motivated by the
results of Gibbons and Hull \cite{gibhul}. The metrics found in \cite{tod} are
bosonic solutions of minimal $N=2$ supergravity theory admitting half of the
supersymmetry. In the context of the supergravity theory, the Killing spinor
equation represents the vanishing of the gravitini supersymmetry
transformation in a bosonic background. The metrics with a time-like Killing
vector are the known Israel-Wilson-Perj\'{e}s (IWP) metrics \cite{IWP} with
the static limit given by the Majumdar-Papapetrou (MP) metrics \cite{mp}. The
second class of metrics with a null-Killing vector is given by plane-wave
space-times \cite{book}. In recent years, a considerable amount of research
activities has been devoted to the understanding and the systematic
classification of supersymmetric solutions in ungauged, gauged and fake (de
Sitter) supergravity theories in various dimensions (see for example
\cite{recentlower}). Fake de Sitter supergravity can be obtained by analytic
continuation of anti de Sitter supergravity. We also note that de Sitter
supergravities can also be obtained as genuine low energy effective theories
of the so called * theories of \cite{hull}. For instance, a non-linear Kaluza
Klein reduction arising of IIB* string theory and M* theory produce four and
five-dimensional de Sitter supergravities with vector multiplets. However
these theories have actions where some of the gauge fields kinetic terms have
the non-conventional sign \cite{jimwen}. Black hole solutions with anti or
phantom Maxwell fields\footnote{where anti or phantom Maxwell field refers to
an electromagnetic field of the opposite sign from usual} have been studied
and analyzed in \cite{dyson}. Black hole solutions with phantom fields and
their relations to astrophysics and dark matter were also considered by many
authors (see \cite{comref} and references therein). However, to our knowledge
phantom solutions with Killing spinors have not yet been discussed.

In our present work, we shall study metrics admitting Killing spinors in
gravitational theories with anti-Maxwell fields. We shall only focus on the
simplest theory of four-dimensional Einstein gravity coupled to a Maxwell
field as a first step for a future study of supergravity theories with many
anti-Maxwell and scalar fields in various space-time dimensions. We will
consider both the Lorentzian and the Euclidean theory. The action of the
theory is given by
\begin{equation}
S=\int\ \mathrm{d}^{4}x\sqrt{-g}\ \left(  R+\kappa^{2}F_{\mu\nu}F^{\mu\nu
}\right)  ,\label{ac}%
\end{equation}
where $F_{\mu\nu}$ is the $U(1)$ gauge field strength. We have introduced a
parameter $\kappa$ which for $\kappa=i,$ corresponds to the standard
Einstein-Maxwell theory and for $\kappa=1$ corresponds to the
Einstein-anti-Maxwell theory, i.e., where the Maxwell field kinetic term comes
with the wrong sign. The signature of the metric is taken to be $(-,+,+,+)$.
For $\kappa=1,$ this action can be thought of as the bosonic part of a fake
minimal $N=2,$ $D=4$ supergravity. The Einstein and gauge field equations
derived from (\ref{ac}) are
\begin{align}
R_{\mu\nu} &  =-\kappa^{2}\left(  2F_{\mu\rho}F_{\nu}{}^{\rho}-{\frac{1}{2}%
}g_{\mu\nu}F_{\alpha\beta}F^{\alpha\beta}\right)  ,\nonumber\\
d\ast F &  =0\ .
\end{align}
Here $F$ is the two form representing the gauge field strength $F_{\mu\nu}.$
The Killing spinor equation is given by
\begin{equation}
\left(  \partial_{\mu}+{\frac{1}{4}}\omega_{\mu,\nu_{1}\nu_{2}}\gamma^{\nu
_{1}\nu_{2}}+{\frac{\kappa}{4}}F_{\nu_{1}\nu_{2}}\gamma^{\nu_{1}\nu_{2}}%
\gamma_{\mu}\right)  \varepsilon=0.\label{killing}%
\end{equation}
where $\varepsilon$ is a non-zero Dirac Killing spinor and $\omega_{\mu
,\nu_{1}\nu_{2}}$ are the spin connections components. For $\kappa=i,$ the
Killing spinor equation is simply the vanishing of the gravitini supersymmetry
transformation in a bosonic background of minimal $N=2,$ $D=4$ supergravity.

We note that if one interchanges $F$ by its Hodge dual $\ast$ $F$ in the
Lorentzian Einstein--Maxwell equations, this simply maps solutions to
solutions as Lorentzian Maxwell stress energy tensor is unchanged by this
transformation. In the Euclidean case, however, this is no longer the case and
the stress energy tensor picks up a minus sign. Therefore with Euclidean
signatures, solutions for theory with the wrong sign of the coupling of
Maxwell field are those for the theory with the "correct" sign of the coupling
but with $F$ $\ $and $\ast$ $F$ interchanged \cite{cosomoinstantons}. This can
also be seen from the inspection of the Killing spinor equations
\cite{instanton}.

We shall use the spinorial geometry method which has proved to be a very
powerful method in the classification of geometric backgrounds admitting
various fractions of supersymmetry in supergravity theories. The isomorphism
between Clifford algebras and exterior algebras allows one to express the
Killing spinor in terms of differential forms. The canonical forms of the
spinor are basically representatives up to gauge transformations which
preserve the supercovariant connection (the reader can refer to \cite{spin}
for spin geometry as well as supersymmetric black holes classifications).

Following \cite{spin}, Dirac spinors in four space-time dimensions can be
written as complexified forms on $\mathbb{R}^{2}$%

\begin{equation}
\varepsilon=\lambda1+\mu_{1}e^{1}+\mu_{2}e^{2}+\sigma e^{12}, \label{gen}%
\end{equation}
where $e^{1}$, $e^{2}$ are 1-forms on $\mathbb{R}^{2}$ and $e^{12}=e^{1}\wedge
e^{2}$. The functions $\lambda$, $\mu_{i}$ and $\sigma$ are complex functions.
The action of $\gamma$-matrices on these forms is given by%

\begin{align}
\gamma_{0}  &  =-e^{2}\wedge+i_{e^{2}},\text{ \ \ \ }\nonumber\\
\gamma_{1}  &  =e^{1}\wedge+i_{e^{1}},\nonumber\\
\gamma_{2}  &  =e^{2}\wedge+i_{e^{2}},\text{ \ \ \ \ \ \ }\nonumber\\
\gamma_{3}  &  =i(e^{1}\wedge-i_{e^{1}})\ .
\end{align}
and $\gamma_{5}$ is defined by $\gamma_{5}=i\gamma_{0123}$ and satisfies%

\begin{equation}
\gamma_{5}1={1},\text{ \ \ \ \ \ \ \ \ }\gamma_{5}e^{12}=e^{12},\text{
\ \ \ \ \ \ \ }\gamma_{5}e^{i}=-e^{i},\text{ }i=1,2.
\end{equation}
Following \cite{gutsabra} we define%

\begin{align}
\gamma_{+}  &  ={\frac{1}{\sqrt{2}}}(\gamma_{2}+\gamma_{0})=\sqrt{2}i_{e^{2}%
},\text{ \ \ \ \ \ \ \ \ }\nonumber\\
\gamma_{-}  &  ={\frac{1}{\sqrt{2}}}(\gamma_{2}-\gamma_{0})=\sqrt{2}%
e^{2}\wedge,\nonumber\\
\gamma_{1}  &  ={\frac{1}{\sqrt{2}}}(\gamma_{1}+i\gamma_{3})=\sqrt{2}i_{e^{1}%
},\text{ \ \ \ \ \ \ \ }\nonumber\\
\gamma_{\bar{1}}  &  ={\frac{1}{\sqrt{2}}}(\gamma_{1}-i\gamma_{3})=\sqrt
{2}e^{1}\wedge. \label{actions}%
\end{align}
In this basis the non-zero metric components are given by $g_{+-}%
=1,g_{1\bar{1}}=1$.\ 

As has been demonstrated in \cite{gutsabra}, using $Spin(3,1)$ gauge
transformations, one finds the three canonical orbits:
\begin{equation}
\text{\ \ \ }\varepsilon=1+\mu_{2}e^{2},\text{ \ \ \ }\varepsilon=1+\mu
_{1}e^{1},\text{ \ \ \ }\varepsilon=e^{2},\label{orbits}%
\end{equation}
where $\mu_{1}$ and $\mu_{2}$ are complex function. Note that the first orbit
represents the Killing spinor for the IWP\ metric which has a time-like
Killing vector. The other two orbits correspond to plane-waves with null
Killing vector. In this letter we are interested in finding phantom solutions
for the Killing spinor \ $\varepsilon=1+\mu e^{2}.$ Similarly we consider the
analogue solutions in the Euclidean case.

\section{Phantom IWP solutions}

For the orbit $\varepsilon=1+\mu e^{2}$, the integrability conditions of the
Killing spinor equation are consistent with the equations of motion. Any
solution of the Killing spinor equation in which the gauge field satisfies the
Bianchi identity and Maxwell equation is automatically a solution of Einstein
equations of motion.

Our solution can be written in the form%

\begin{equation}
ds_{4}^{2}=2\mathbf{e}^{+}\mathbf{e}^{-}+2\mathbf{e}^{1}\mathbf{e}^{\bar{1}}.
\end{equation}
Plugging $\varepsilon=1+\mu e^{2}$ in (\ref{killing}) and using (\ref{actions}%
), the Killing spinor equations amounts to a set of sixteen algebraic and
differential equations:
\begin{align}
-(\omega_{+,+-}+\omega_{+,1\bar{1}})-\sqrt{2}\kappa\mu(F_{+-}+F_{1\bar{1}})
&  =0,\nonumber\\
\omega_{+,-1}  &  =0,\nonumber\\
\partial_{+}\mu+\frac{\mu}{2}(\omega_{+,+-}-\omega_{+,1\bar{1}})  &
=0,\nonumber\\
\omega_{+,+1}+\kappa\sqrt{2}\mu F_{+1}  &  =0,\nonumber\\
\omega_{-,+-}+\omega_{-,1\bar{1}}  &  =0,\nonumber\\
\mu\omega_{-,-1}+\kappa\sqrt{2}F_{-1}  &  =0,\nonumber\\
\partial_{-}\mu+\frac{\mu}{2}(\omega_{-,+-}-\omega_{-,1\bar{1}})+\frac{\kappa
}{\sqrt{2}}(F_{+-}-F_{1\bar{1}})  &  =0,\nonumber\\
\omega_{-,+1}  &  =0,\nonumber\\
\omega_{1,+-}+\omega_{1,1\bar{1}}  &  =0,\nonumber\\
\omega_{1,-1}  &  =0,\nonumber\\
\partial_{1}\mu+\frac{\mu}{2}(\omega_{1,+-}-\omega_{1,1\overline{1}})  &
=0,\nonumber\\
\omega_{1,+1}  &  =0,\nonumber\\
-\frac{1}{2}(\omega_{\overline{1},+-}+\omega_{\overline{1},1\overline{1}%
})+\kappa\sqrt{2}\mu F_{-\overline{1}}  &  =0,\nonumber\\
\mu\omega_{\overline{1},-1}+\frac{\kappa}{\sqrt{2}}(F_{+-}-F_{1\overline{1}})
&  =0,\nonumber\\
\partial_{\bar{1}}\mu+\frac{\mu}{2}(\omega_{\bar{1},+-}-\omega_{\bar{1}%
,1\bar{1}})+\kappa\sqrt{2}F_{+\bar{1}}  &  =0,\nonumber\\
\omega_{\bar{1},+1}-\frac{\kappa}{\sqrt{2}}\mu(F_{+-}+F_{1\bar{1}})  &  =0.
\end{align}

The analysis of this system of equations gives the following relations for the
gauge field strength components
\begin{align}
F_{+-}  &  =-\frac{1}{\sqrt{2}}\partial_{-}\left(  \kappa\bar{\mu}+\bar
{\kappa}\mu\right)  ,\text{ \ \ }\nonumber\\
F_{1\overline{1}}  &  =-\frac{1}{\sqrt{2}}\partial_{-}\left(  \kappa\bar{\mu
}-\bar{\kappa}\mu\right)  ,\nonumber\\
\text{ }F_{-1}  &  =-\frac{\kappa}{\sqrt{2}|\mu|^{2}}\partial_{1}\mu,\text{
\ \ \ \ \ }\nonumber\\
F_{+1}  &  =-\frac{\kappa}{\sqrt{2}}\partial_{1}\bar{\mu}, \label{gfs}%
\end{align}
together with the relation%
\begin{equation}
\left(  \partial_{+}+\kappa^{2}|\mu|^{2}\partial_{-}\right)  \mu=0.
\label{conderivatives}%
\end{equation}
We also obtain for the spin connections%

\begin{align}
\omega_{+-} &  =\kappa^{2}\partial_{-}|\mu|^{2}\mathbf{e}^{+}-\partial_{1}%
\log\mu\mathbf{e}^{1}-\partial_{\bar{1}}\log\bar{\mu}\mathbf{e}^{\bar{1}%
},\text{ \ \ }\nonumber\\
\omega_{1\overline{1}} &  =\kappa^{2}|\mu|^{2}\partial_{-}\log\frac{\bar{\mu}%
}{\mu}\mathbf{e}^{+}+\partial_{1}\log\mu\mathbf{e}^{1}-\partial_{\bar{1}}%
\log\bar{\mu}\mathbf{e}^{\bar{1}},\nonumber\\
\omega_{+1} &  =-\kappa^{2}\mu\left(  \partial_{-}\bar{\mu}\mathbf{e}^{\bar
{1}}-\partial_{1}\bar{\mu}\mathbf{e}^{+}\right)  ,\nonumber\\
\omega_{-1} &  =\frac{1}{\mu}\left(  \frac{\kappa^{2}}{|\mu|^{2}}\partial
_{1}\mu\mathbf{e}^{-}+\partial_{-}\mu\mathbf{e}^{\bar{1}}\right)  .\label{geo}%
\end{align}
The equations (\ref{conderivatives}) and (\ref{geo}) can be used to
demonstrate that the vector $V,$%

\begin{equation}
V=|\mu|^{2}\mathbf{e}^{+}+\kappa^{2}\mathbf{e}^{-}=|\mu|^{2}\partial
_{-}+\kappa^{2}\partial_{+} \label{kilvec}%
\end{equation}
is a Killing vector which is space-like for $\kappa^{2}=1$ and time-like for
$\kappa^{2}=-1.$

Moreover, the vanishing of the torsion, i.e.,
\begin{equation}
d\mathbf{e}^{a}+\omega_{\text{ }b}^{a}\wedge\mathbf{e}^{b}=0,
\end{equation}
implies the following relations%

\begin{equation}
d\mathbf{e}^{1}=-d\left(  \log\bar{\mu}\right)  \wedge\mathbf{e}^{1},
\label{done}%
\end{equation}
and%

\begin{equation}
d\mathbf{e}^{+}=\left(  \mathbf{e}^{+}-\frac{\kappa^{2}}{|\mu|^{2}}%
\mathbf{e}^{-}\right)  \wedge\left(  \partial_{1}\log\mu\mathbf{e}%
^{1}+\partial_{\bar{1}}\log\bar{\mu}\mathbf{e}^{\bar{1}}\right)  +\partial
_{-}\log\frac{\mu}{\bar{\mu}}\mathbf{e}^{1}\wedge\mathbf{e}^{\bar{1}},
\label{deplus}%
\end{equation}
and%

\begin{align}
d\mathbf{e}^{-}  &  =-\kappa^{2}\partial_{-}\left(  |\mu|^{2}\right)
\mathbf{e}^{+}\wedge\mathbf{e}^{-}+\left(  \partial_{1}\log\mu\mathbf{e}%
^{1}+\partial_{\bar{1}}\log\bar{\mu}\mathbf{e}^{\bar{1}}\right)
\wedge\mathbf{e}^{-}\nonumber\\
&  -\kappa^{2}\left(  \mu\partial_{-}\bar{\mu}-\bar{\mu}\partial_{-}%
\mu\right)  \mathbf{e}^{1}\wedge\mathbf{e}^{\bar{1}}-\kappa^{2}\mathbf{e}%
^{+}\wedge\left(  \bar{\mu}\partial_{\bar{1}}\mu\mathbf{e}^{\bar{1}}%
+\mu\partial_{1}\bar{\mu}\mathbf{e}^{1}\right)  . \label{deminus}%
\end{align}
Using these relations, it can be shown that $\left(  |\mu|^{2}\mathbf{e}%
^{+}-\kappa^{2}\mathbf{e}^{-}\right)  $ satisfies%

\begin{equation}
d\left(  |\mu|^{2}\mathbf{e}^{+}-\kappa^{2}\mathbf{e}^{-}\right)  =0.
\label{differential}%
\end{equation}
\ and thus is a total differential\footnote{Note that $\left(  |\mu
|^{2}\mathbf{e}^{+}+\kappa^{2}\mathbf{e}^{-}\right)  $ and $|\left(  \mu
|^{2}\mathbf{e}^{+}-\kappa^{2}\mathbf{e}^{-}\right)  $ are related to
Hermitian inner products by
\par
{}%
\[
\sqrt{2}\left(  |\mu|^{2}\mathbf{e}^{+}-\mathbf{e}^{-}\right)  =<\gamma
_{0}\varepsilon,\gamma_{a}\varepsilon>;\ \ \ \ \ \ \ \sqrt{2}\left(  |\mu
|^{2}\mathbf{e}^{+}+\mathbf{e}^{-}\right)  =<\gamma_{0}\varepsilon,\gamma
_{5}\gamma_{a}\varepsilon>.
\]
}.

The relations (\ref{kilvec}), (\ref{done}) and (\ref{differential}), enable us
to introduce the real coordinates $\left(  t,x,y,z\right)  $ coordinates, such
that
\begin{align}
\mathbf{e}^{-}  &  =-\frac{1}{\sqrt{2}}\left(  \kappa^{2}dz-|\mu|^{2}\left(
dt+\phi\right)  \right)  ,\text{ \ \ \ \ \ }\nonumber\\
\mathbf{e}^{+}  &  =\frac{1}{\sqrt{2}|\mu|^{2}}\left(  dz+\kappa^{2}|\mu
|^{2}\left(  dt+\phi\right)  \right)  ,\nonumber\\
\mathbf{e}^{1}  &  =\frac{1}{\bar{\mu}\sqrt{2}}\left(  dx+idy\right)  .
\label{viel}%
\end{align}

Here $\phi$ is a one form independent of the coordinate $t$ and $\phi_{t}=0.$
\ Note that (\ref{conderivatives}) implies that $\mu$ is independent of $t.$
The metric is therefore given by%

\begin{equation}
ds_{4}^{2}=2\mathbf{e}^{+}\mathbf{e}^{-}+2\mathbf{e}^{1}\mathbf{e}^{\bar{1}%
}=\kappa^{2}|\mu|^{2}\left(  dt+\phi\right)  ^{2}+\frac{1}{|\mu|^{2}}%
ds_{3}^{2}%
\end{equation}
where $ds_{3}^{2}=\left(  -\kappa^{2}dz^{2}+dx^{2}+dy^{2}\right)  .$ Moreover,
substituting the relations (\ref{viel}) into (\ref{deplus}) or (\ref{deminus}%
), one can derive the relation%

\begin{equation}
d\phi=\frac{i\kappa^{2}}{|\mu|^{2}}\ast_{3}d\log\frac{\bar{\mu}}{\mu
},\label{dsigma}%
\end{equation}
where $\ast_{3}$ is the Hodge dual with metric $ds_{3}^{2}.$

Using (\ref{gfs}), (\ref{viel}) and%

\begin{equation}
\partial_{+}=\frac{|\mu|^{2}}{\sqrt{2}}\partial_{z},\text{ \ \ \ \ \ }%
\partial_{-}=-\frac{\kappa^{2}}{\sqrt{2}}\partial_{z},\text{ \ \ \ \ }%
\partial_{1}=\frac{\bar{\mu}}{\sqrt{2}}\left(  \partial_{x}-i\partial
_{y}\right)  , \label{der}%
\end{equation}
we obtain for the gauge field strength two form%

\begin{equation}
F=\frac{1}{2}d\left(  \bar{\kappa}\bar{\mu}+\kappa\mu\right)  \wedge\left(
dt+\phi\right)  -\frac{i}{2|\mu|^{2}}\ast_{3}d\left(  \bar{\kappa}\mu
-\kappa\bar{\mu}\right)  .
\end{equation}
The dual gauge field strength two form is given by%

\[
\ast\text{ }F=\frac{i}{2}d\left(  \bar{\kappa}\bar{\mu}-\kappa\mu\right)
\wedge\left(  dt+\phi\right)  -\frac{1}{2|\mu|^{2}}\ast d\left(  \bar{\kappa
}\mu+\kappa\bar{\mu}\right)  .
\]
Using (\ref{dsigma}), $F$ and $\ast$ $F$ can be rewritten in the form%

\begin{align}
F &  =\frac{1}{2}d\left[  \left(  \frac{\bar{\mu}}{\kappa}+\frac{\mu}%
{\bar{\kappa}}\right)  \left(  dt+\phi\right)  \right]  -\frac{i}{2}\ast
d\left(  \frac{\kappa}{\mu}-\frac{\bar{\kappa}}{\bar{\mu}}\right)
,\nonumber\\
\ast F &  =\frac{i}{2}d\left[  \left(  \frac{\mu}{\kappa}-\frac{\bar{\mu}%
}{\bar{\kappa}}\right)  \left(  dt+\phi\right)  \right]  -\frac{1}{2}\ast
d\left(  \frac{\kappa}{\mu}+\frac{\bar{\kappa}}{\bar{\mu}}\right)  .
\end{align}
Then Bianchi identity together with Maxwell equation imply that
\begin{align}
\nabla^{2}\left(  \frac{\kappa}{\mu}-\frac{\bar{\kappa}}{\bar{\mu}}\right)
&  =\nabla^{2}\left(  \frac{\kappa}{\mu}+\frac{\bar{\kappa}}{\bar{\mu}%
}\right)  =0,\nonumber\\
\nabla^{2} &  =\left(  \partial_{x}^{2}+\partial_{y}^{2}-\kappa^{2}%
\partial_{z}^{2}\right)  .
\end{align}
For $\kappa=i,$ the solution obtained is the IWP\ metric \cite{IWP} where the
inverse of $\mu$ is a complex harmonic function . For $\kappa=1,$ we obtain
the new solutions in which the inverse of $\mu$ satisfies the wave equation in
flat $(2+1)$-space-time. For $\kappa=1,$ $\mu=\bar{\mu},$ we obtain the
analogue of the electric MP solution \cite{mp}%
\begin{align}
ds^{2} &  =\mu^{2}dt^{2}+\frac{1}{\mu^{2}}\left(  -dz^{2}+dx^{2}%
+dy^{2}\right)  ,\nonumber\\
A &  =\mu dt,\nonumber\\
\left(  \partial_{x}^{2}+\partial_{y}^{2}-\kappa^{2}\partial_{z}^{2}\right)
\left(  \frac{1}{\mu}\right)   &  =0.
\end{align}
where $A$ is the gauge field one form. For $\mu=i\alpha,$ we get the magnetic
solution%
\begin{align}
ds^{2} &  =\alpha^{2}dt^{2}+\frac{1}{\alpha^{2}}\left(  -dz^{2}+dx^{2}%
+dy^{2}\right)  \text{ }\nonumber\\
F &  =-\ast_{3}d\left(  \frac{1}{\alpha}\right)  ,\nonumber\\
\left(  \partial_{x}^{2}+\partial_{y}^{2}-\kappa^{2}\partial_{z}^{2}\right)
\left(  \frac{1}{\alpha}\right)   &  =0.
\end{align}

\subsection{Charged Kasner Universe}

As special interesting phantom metric examples, we take as a solution to the
wave equation in flat $(2+1)$-space-time,
\begin{equation}
\mu=\frac{q}{z}%
\end{equation}
with constant $q.$ The metric and the gauge field strength then take the form%

\begin{align}
ds^{2}  &  =\frac{q^{2}}{z^{2}}dt^{2}-\frac{z^{2}}{q^{2}}dz^{2}+\frac{z^{2}%
}{q^{2}}\left(  dx^{2}+dy^{2}\right)  ,\nonumber\\
F  &  =dA=-\frac{q}{z^{2}}dz\wedge dt.\text{ \ \ \ \ \ \ \ \ \ }%
\end{align}
Introducing the new coordinates%

\begin{equation}
\tau=\frac{1}{2q}z^{2},\text{ \ \ \ \ \ }x^{1}=\sqrt{\frac{2}{q}}x,\text{
\ \ \ \ \ }x^{2}=\sqrt{\frac{2}{q}}y,\text{ \ \ \ \ }x^{3}=\sqrt{\frac{q}{2}%
}t,\text{ \ \ }%
\end{equation}
then the metric takes the Kasner form \cite{kasner}%

\begin{align}
ds^{2}  &  =-d\tau^{2}+\sum_{j=1}^{3}\tau^{2p_{j}}\left(  dx^{j}\right)
^{2},\nonumber\\
F  &  =-\frac{1}{2\tau^{3/2}}d\tau\wedge dx^{3},
\end{align}
with the Kasner exponents%

\begin{equation}
p_{1}=p_{2}=\frac{1}{2},\text{ \ \ \ }p_{3}=-\frac{1}{2}.
\end{equation}
Note that here the Kasner exponents satisfy the conditions%
\begin{equation}
\sum_{j=1}^{3}p_{j}=\frac{1}{2},\text{ \ \ \ \ }\sum_{j=1}^{3}p_{j}^{2}%
=\frac{3}{4},
\end{equation}
while in vacuum they satisfy%

\begin{equation}
\sum_{j=1}^{3}p_{j}=\sum_{j=1}^{3}p_{j}^{2}=1.
\end{equation}
For $\mu=i\alpha=\frac{i}{pz},$ we get the solution%

\begin{align}
ds^{2}  &  =\left(  \frac{1}{pz}\right)  ^{2}dt^{2}+\left(  pz\right)
^{2}\left(  -dz^{2}+dx^{2}+dy^{2}\right)  ,\nonumber\\
F  &  =pdx\wedge dy.
\end{align}
This metric takes the Kasner form%

\begin{align}
ds^{2} &  =-d\tau^{2}+\tau\left(  dx^{1}\right)  ^{2}+\tau\left(
dx^{2}\right)  ^{2}+\tau^{-2}\left(  dx^{3}\right)  ^{2},\nonumber\\
F &  =dx^{1}\wedge dx^{2},
\end{align}
where we have introduced the coordinates%

\begin{equation}
\tau=\frac{1}{2}pz^{2},\text{ \ \ \ }x^{1}=\sqrt{2p}x,\text{ \ \ \ \ \ }%
x^{2}=\sqrt{2p}y,\text{ \ \ \ \ }x^{3}=\sqrt{\frac{1}{2p}}t.
\end{equation}

\section{Euclidean solutions}

As already mentioned, one does not get new exotic solutions with phantom
Euclidean Maxwell fields. The metric solution is independent of the sign of
the coupling of the Maxwell field. For the sake of completeness, we briefly
present the solutions for both couplings in a unified fashion. We take the
metric to be of the form \cite{instanton}%

\begin{equation}
ds^{2}=2\mathbf{e}^{1}\mathbf{e}^{\bar{1}}+2\mathbf{e}^{2}\mathbf{e}^{\bar{2}%
}.
\end{equation}
Dirac spinor is taken to be a linear combination of the complexified space of
forms on $\mathbb{R}^{2}$, with basis $\{1,e_{1},e_{2},e_{12}=e_{1}\wedge
e_{2}\}$. In this basis, the action of the Dirac matrices $\gamma_{m}$ on the
Dirac spinors is given by
\begin{equation}
\gamma_{m}=\sqrt{2}i_{e_{m}},\qquad\gamma_{\bar{m}}=\sqrt{2}e_{m}\wedge
\end{equation}
for $m=1,2$. We also define $\gamma_{5}=\gamma_{1\bar{1}2\bar{2}}$. Euclidean
version of the IWP\ metric were found in \cite{instanton} for the case
$\kappa=i$, and orbit
\begin{equation}
\epsilon=\lambda1+\sigma e_{1}, \label{orb}%
\end{equation}
with real $\lambda$ and $\sigma$. Keeping $\kappa$ as a parameter, then the
analysis of the Killing spinor equation for the orbit (\ref{orb}) gives the
geometric conditions%

\begin{align}
\omega_{1\bar{1}}  &  =\partial_{2}\log\frac{\lambda}{\sigma}\mathbf{e}%
^{2}-\partial_{1}\log\sigma\lambda\mathbf{e}^{1}-\partial_{\bar{2}}\log
\frac{\lambda}{\sigma}\mathbf{e}^{\bar{2}}+\partial_{\bar{1}}\log\sigma
\lambda\mathbf{e}^{\bar{1}}\text{\ },\nonumber\\
\omega_{2\bar{2}}  &  =\partial_{2}\log\lambda\sigma\mathbf{e}^{2}%
-\partial_{\bar{2}}\log\lambda\sigma\mathbf{e}^{\bar{2}}+\partial_{1}\log
\frac{\sigma}{\lambda}\mathbf{e}^{1}-\partial_{\bar{1}}\log\frac{\sigma
}{\lambda}\mathbf{e}^{\bar{1}},\nonumber\\
\omega_{21}  &  =-2\kappa^{2}\partial_{2}\log\lambda\mathbf{e}^{1}%
-2\partial_{1}\log\lambda\mathbf{e}^{\bar{2}},\nonumber\\
\omega_{\bar{2}1}  &  =-2\kappa^{2}\partial_{\bar{2}}\log\sigma\mathbf{e}%
^{1}-2\partial_{1}\log\sigma\mathbf{e}^{2}, \label{geometry}%
\end{align}
together with the condition
\begin{equation}
\left(  \partial_{1}+\kappa^{2}\partial_{\bar{1}}\right)  \sigma=\left(
\partial_{1}+\kappa^{2}\partial_{\bar{1}}\right)  \lambda=0. \label{geometry2}%
\end{equation}
For the gauge field strength we get%

\begin{align}
F_{2\bar{2}} &  =\frac{1}{\sqrt{2}\lambda\sigma}\left[  \partial_{1}\left(
\frac{\lambda^{2}}{\kappa}+\frac{\sigma^{2}}{\bar{\kappa}}\right)
\mathbf{e}^{1}+\text{\ }\frac{\partial_{2}\sigma^{2}}{\bar{\kappa}}%
\mathbf{e}^{2}+\frac{\partial_{\bar{2}}\lambda^{2}}{\kappa}\mathbf{e}^{\bar
{2}}\right]  \wedge\mathbf{e}^{\bar{1}}\ \nonumber\\
&  +\frac{1}{\sqrt{2}\lambda\sigma}\left[  \frac{\partial_{2}\lambda^{2}}%
{\bar{\kappa}}\mathbf{e}^{2}\ +\text{\ }\frac{\partial_{\bar{2}}\sigma^{2}%
}{\kappa}\mathbf{e}^{\bar{2}}\right]  \wedge\mathbf{e}^{1}+\frac{1}{\sqrt
{2}\lambda\sigma}\partial_{1}\left(  \frac{\lambda^{2}}{\kappa}-\frac
{\sigma^{2}}{\bar{\kappa}}\right)  \mathbf{e}^{2}\wedge\mathbf{e}^{\bar{2}}.
\end{align}
For torsion free metric, the conditions (\ref{geometry}) and (\ref{geometry2})
imply that $\lambda\sigma\left(  \mathbf{e}^{1}-\kappa^{2}\mathbf{e}^{\bar{1}%
}\right)  $ is a total differential and that $\kappa\lambda\sigma\left(
\mathbf{e}^{1}+\kappa^{2}\mathbf{e}^{\bar{1}}\right)  $ is a Killing vector.
This enables us to introduce the coordinates $\left(  \tau,x,y,z\right)  $ and write%

\[
\mathbf{e}^{1}=\frac{1}{\sqrt{2}}\left(  -i\kappa\frac{dx}{\lambda\sigma
}+\frac{1}{\kappa}\lambda\sigma\left(  d\tau+\phi\right)  \right)  ,\text{
\ \ }\mathbf{e}^{2}=\frac{1}{\sqrt{2}\lambda\sigma}\left(  dy+idz\right)  ,
\]
and the solution is given by%

\begin{align}
ds^{2}  &  =\left(  \lambda\sigma\right)  ^{2}\left(  d\tau+\phi\right)
^{2}+\frac{1}{\left(  \lambda\sigma\right)  ^{2}}\left(  dx^{2}+dy^{2}%
+dz^{2}\right)  ,\nonumber\\
d\phi &  =\frac{2\kappa^{2}}{\left(  \lambda\sigma\right)  ^{2}}\ast
d\log\frac{\lambda}{\sigma},\nonumber\\
F  &  =\frac{1}{2}d\left[  \left(  \kappa^{2}\sigma^{2}+\lambda^{2}\right)
\left(  d\tau+\phi\right)  \right]  +\frac{1}{2}\ast d\left(  \frac{1}%
{\lambda^{2}}-\frac{\kappa^{2}}{\sigma^{2}}\right)  ,\nonumber\\
\tilde{F}  &  =-\frac{1}{2}d\left[  \left(  \lambda^{2}-\kappa^{2}\sigma
^{2}\right)  \left(  d\tau+\phi\right)  \right]  +\frac{1}{2}\ast d\left(
\frac{1}{\lambda^{2}}+\frac{\kappa^{2}}{\sigma^{2}}\right)  .
\end{align}
with $\lambda$ and $\sigma$ independent of $\tau.$ The Bianchi identity and
Maxwell equation imply the equations
\begin{equation}
\nabla^{2}\left(  \frac{1}{\lambda^{2}}-\frac{\kappa^{2}}{\sigma^{2}}\right)
=\nabla^{2}\left(  \frac{1}{\lambda^{2}}+\frac{\kappa^{2}}{\sigma^{2}}\right)
=0,
\end{equation}
where $\nabla^{2}=\partial_{x}^{2}+\partial_{y}^{2}+\partial_{z}^{2}.$

\bigskip

In summary, the method of spinorial geometry is used to find IWP\ analogue
solutions in four-dimensional Einstein-anti-Maxwell theory admitting Killing
spinors. The analysis of the Killing spinor equation reveals the existence of
a Killing vector and a total differential which switch roles when one changes
the coupling of the Maxwell field. The phantom solutions found admit a
space-like Killing vector and constitute the time-dependent analogues of the
IWP metrics of the canonical Einstein-Maxwell theory. The solutions are
expressed in terms of a complex function satisfying the wave equation in a
flat $(2+1)$-space-time. As examples, electric and magnetic Kasner spaces can
be constructed by specializing to solutions that depend only on the time
coordinate. The Kasner exponent sum rules of the vacuum Kasner solution get
modified in the presence of a phantom $U(1)$ gauge field. Phantom Euclidean
solutions are also presented. In the Euclidean case, the phantom metric is the
same as in the ordinary Einstein-Maxwell theory but with the roles of $F$ and
$\ast$ $F$ interchanged. Our analysis can be extended to theories with
anti-scalars and anti-vector multiplets in ungauged and gauged supergravity
models in various dimensions. Work in this direction is in progress.

\bigskip

\textbf{Acknowledgements} : \ The work of W. S is supported in part by the
National Science Foundation under grant number PHY-1415659. The author would
like to thank D. Klemm and G. W. Gibbons for useful discussions.

\bigskip
\end{document}